\documentclass[twocolumn,showpacs,prb,aps,amsmath,amssymb,superscriptaddress]{revtex4}

\usepackage{graphicx}
\usepackage{bm}
\usepackage{color}

\begin{document}

\title{
Sine-square deformation of free fermion systems in one and higher dimensions
}
\author{Isao Maruyama}
\affiliation{Graduate School of Engineering Science, Osaka University, Toyonaka, Osaka 560-8531, Japan}
\author{Hosho Katsura}
\affiliation{Department of Physics, Gakushuin University, Toshima-ku, Tokyo 171-8588, Japan}
\author{Toshiya Hikihara}
\affiliation{Faculty of Engineering, Gunma University, Kiryu, Gunma 376-8515, Japan}

\date{\today}

\begin{abstract}
We study free fermion systems with the sine-square deformation (SSD), in which the energy scale of local Hamiltonians is modified according to the scaling function $f(x)=\sin^2\left[\frac{\pi}{L}(x-\frac{1}{2})\right]$, where $x$ is the position of the local Hamiltonian and $L$ is the length of the system in the $x$ direction. It has been revealed that when applied to one-dimensional critical systems the SSD realizes the translationally-invariant ground state which is the same as that of the uniform periodic system. In this paper, we propose a simple theory to explain how the SSD maintains the translational invariance in the ground-state wave function. In particular, for a certain one-dimensional system with SSD, it is shown that the ground state is exactly identical with the Fermi sea of the uniform periodic chain. We also apply the SSD to two-dimensional systems and show that the SSD is able to suppress the boundary modulations from the open edges extremely well, demonstrating that the SSD works in any dimensions and in any directions.

\end{abstract}


\maketitle

\section{Introduction}\label{sec:intro}

The boundary condition has a crucial influence on properties of quantum systems.
For example, the periodic boundary condition (PBC) realizes a translationally-invariant system with the geometry of the ring or torus, while the open boundary condition (OBC) results in the chain or cylinder with open boundaries with the boundary-induced modulation such as the Friedel oscillations.
The thermodynamic properties, i.e., the expectation values of observables in the bulk of infinite-size systems, are believed to be the same irrespective of the boundary conditions employed.
However, the way that the expectation values approach the thermodynamic limit is usually distinct for OBC and PBC, which requires a different treatment of finite-size scaling.
Furthermore, at the level of the wave function, the difference between the critical systems under OBC and PBC remains sizable even at the thermodynamic limit. That is clearly demonstrated by the different scaling behavior of the entanglement entropy in one-dimensional (1d) critical systems with and without open boundaries.\cite{Calabrese2004}

Attempts to control the boundary and finite-size effects have been made in the studies of lattice fermion/spin models.
One example is the so-called smooth boundary condition, which succeeded pretty well in suppressing the boundary effects by smoothly decreasing the energy scales near the edges of the system.\cite{VekicW1993,VekicW1996} 
Similar but modified boundary conditions were applied to transport problems such as the conductance of 1d interacting systems.\cite{Bohr05}
Another example is the hyperbolic deformation in which the ground state expectation values of local observables are nearly uniform in the bulk whereas the interaction strength is enhanced rather than suppressed at the boundaries. \cite{Ueda-Nishino08, Ueda-Nakano08, Ueda-Nakano10}

Recently, a more efficient scheme to suppress the boundary effects, which is called the sine-square deformation (SSD), has been proposed for 1d critical systems.
In the system with SSD, the energy scale of the local Hamiltonian is rescaled according to the scaling function,
\begin{eqnarray}
f(x) = \sin^2\left[ \frac{\pi}{L} \left( x - \frac{1}{2} \right) \right],
\label{eq:fx}
\end{eqnarray}
where $L$ is the length of the system and $x$ is the center position of the local Hamiltonian.
Note that since $f(x) = 0$ for $x=1/2$ (mod $L$), the SSD disconnects the link between the sites at $x=1$ and $L$, resulting in the open edges.
The SSD was first applied to the 1d free-fermion system.\cite{GendiarKN2009}
It was found there that in the system with SSD the local observables such as the bond strength are translationally invariant, suggesting that the SSD was able to remove the open-boundary effects almost completely.
Furthermore, the effects of SSD in several quantum spin models at criticality were examined numerically.\cite{HikiharaN2011}
In addition to the translational invariance of the observables such as spin correlations, it was further revealed that the wave-function overlap between the ground state of the open system with SSD and that of the uniform system with PBC is very close to (almost exactly) unity.
This means that the SSD does not disturb the periodic ground state at the level of the wave function and keeps the state as the true ground state even after the model Hamiltonian loses the translational symmetry by the energy deformation.
This striking feature of the SSD was proven rigorously for the spin-1/2 XY chain, which is equivalent to the free fermions in one dimension.\cite{Katsura2011}
It has also been shown that the SSD works well for other strongly correlated systems including the 1d Hubbard model\cite{GendiarDLN2011} and Kondo-lattice model.\cite{ShibataH2011}

While the efficiency of the SSD has been confirmed for several 1d systems, it remains a puzzle why the ground state of the system with SSD is (almost) identical to that of the periodic system.
Since the system with SSD is no longer translational invariant, its one-particle eigenstates are distinct from the plane waves. 
However, when (and only when) the fermions are filled up to the Fermi level, the many-particle ground states of the system with SSD and the uniform periodic system become equivalent.
It is desirable to clarify the mechanism of the SSD to leave the ground state under PBC unchanged.

In this paper, we propose a simple theory which explains how the SSD realizes the energy-scale deformation with keeping the ground state of the original free-fermion model with PBC as an eigenstate. 
The key idea is that the amplitudes of active processes in the Hamiltonian with SSD vanish, therefore the Hamiltonian does not disturb the Fermi sea of the original model with PBC.
For a certain class of 1d systems, it is also proved that the Fermi sea of the original uniform system is not only an eigenstate but also the unique ground state of the Hamiltonian with SSD.
For two-dimensional (2d) systems, the ground state of the system with SSD is found to be very close to the Fermi sea of the original periodic system accompanied by edge states localized around open edges.
We emphasize that the SSD works well to suppress the boundary effect induced by the OBC in any dimensions and in any directions.
The SSD thereby reduces the number of directions under PBC with maintaining the ground-state properties in the bulk.
For example, applying the SSD to a 2d model changes its geometry from a torus to a cylinder, then to a rectangle.

The paper is organized as follows.
In Sec.\ \ref{sec:mechanism}, we discuss the Hamiltonian with SSD in the momentum space and show that the Fermi sea of the uniform system remains an (approximate) eigenstate through the energy deformation.
The results for the 1d and 2d systems are presented in Sec.\ \ref{sec:SSD1d} and \ref{sec:SSD2d}, respectively.
In Sec.\ \ref{sec:numerical}, we study numerically the effects of the SSD on the energy-level structure as well as the ground state properties of 2d systems.
By investigating several quantities such as the one-particle eigen-energies, eigen-wavefunctions and the ground-state density profiles, we examine how the ground state is modified by the SSD.
We conclude with some remarks in Sec.\ \ref{sec:conclusion}.

\section{SSD in momentum space}\label{sec:mechanism}

\subsection{One-dimensional systems}\label{sec:SSD1d}

In this section, we develop a theory of the mechanism of the SSD applied to free spinless fermions on a lattice.
We first focus on the simplest model, i.e., the 1d tight-binding model with only nearest-neighbor hopping.
For the model, it was rigorously proved that the model with SSD shares the same ground state with the model under PBC.\cite{Katsura2011}

The model Hamiltonian of the tight-binding chain is given by
\begin{eqnarray}
{\cal H} &=& -t \sum_x \left[ c^\dagger(x) c(x+1) + {\rm H.c.} \right]
-\mu \sum_x c^\dagger(x) c(x),
\nonumber \\
\label{eq:Ham-chain}
\end{eqnarray}
where $c^\dagger(x)$ and $c(x)$ are respectively the creation and annihilation operators of fermion at the site $x$.
We consider the $L$-site system and impose the PBC.
The Hamiltonian in momentum space is obtained as 
\begin{eqnarray}
\tilde{\cal H} = \sum_k \epsilon(k) c^\dagger_k c_k
\end{eqnarray}
with the dispersion relation
\begin{eqnarray}
\epsilon(k) = -2t \cos k -\mu.
\label{eq:disp-chain}
\end{eqnarray}

A key ingredient of the construction of SSD is to introduce the chiral deformation, which leads to the chiral Hamiltonian of the form,
\begin{eqnarray}
{\cal H}^{(\pm)} &=& -t \sum_x e^{\pm i\delta x} \left[ c^\dagger(x) c(x+1) + {\rm H.c.} \right]
\nonumber \\
&&-\mu \sum_x e^{\pm i\delta (x-\frac{1}{2})} c^\dagger(x) c(x),
\end{eqnarray}
where $\delta = 2\pi/L$. 
The chiral Hamiltonian ${\cal H}^{(\pm)}$ is nothing but the Fourier component of the local Hamiltonians and is apparently non-Hermitian.~\cite{Katsura-Lee-Nagaosa10} 
In the basis where the original Hamiltonian ${\cal H}$ is diagonal, ${\cal H}^{(\pm)}$ is represented as
\begin{eqnarray}
\tilde{\cal H}^{(\pm)} = \sum_k
e^{\mp i \frac{\delta}{2}} \epsilon(k \mp \delta/2)
c^\dagger_k c_{k \mp \delta}.
\label{eq:Fourier-chain}
\end{eqnarray}

\begin{figure}
\begin{center}
\includegraphics[width=65mm]{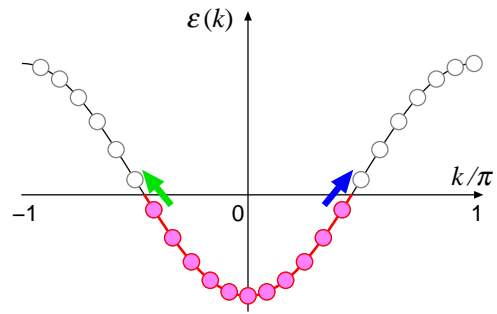}
\caption{
Dispersion curve and Fermi sea of the 1d tight-binding model (\ref{eq:Ham-chain}) with $t > 0$ and $\mu < 0$. Dark (blue) and light (green) arrows represent respectively the processes in the chiral Hamiltonian $\tilde{\cal H}^{(+)}$ and $\tilde{\cal H}^{(-)}$ allowed by the Pauli principle.
}
\label{fig:chain}
\end{center}
\end{figure}

The point to be noted here is that the chiral Hamiltonian includes only the terms with momentum transfer $\Delta k = \pm \delta$, which connects the nearest-neighbor sites in $k$ space, and their amplitude is proportional to the dispersion function Eq.\ (\ref{eq:disp-chain}) of the original uniform system.
Hence, when applied to the Fermi sea of the original model, most of the transfer processes are prohibited by the Pauli principle; 
Only the processes across the Fermi point $k=\pm k_{\rm F}$ can be active.
(The schematic picture of the active processes is shown in Fig.\ \ref{fig:chain}.)
However, for such processes, the amplitudes are, at most, of the order of ${\cal O}(1/L)$.
Indeed, one can make these amplitudes be exactly zero by fine-tuning of the chemical potential $\mu$ so that $\epsilon(k \mp \delta/2) = 0$ for the active processes.
This means that the Fermi sea of the original uniform chain is an exact eigenstate of the chiral Hamiltonian ${\cal H}^{(\pm)}$ and is not disturbed by the chiral deformation at all.
The same result holds for any 1d systems having a single Fermi momentum $k_{\rm F}$ and symmetric dispersion $\epsilon(k)=\epsilon(-k)$.

Now, we are ready to consider the system with SSD.
The Hamiltonian with SSD is constructed from the original and chiral Hamiltonians as
\begin{eqnarray}
{\cal H}_{\rm SSD} &=& 
\frac{1}{2} {\cal H} - \frac{1}{4} \left[ {\cal H}^{(+)} + {\cal H}^{(-)} \right]
\nonumber \\
&=&
-t \sum_x f(x+1/2) \left[ c^\dagger(x) c(x+1) + {\rm H.c.} \right]
\nonumber \\
&&-\mu \sum_x f(x) c^\dagger(x) c(x),
\label{eq:Ham-SSD-chain}
\end{eqnarray}
where the scaling function $f(x)$ is given by Eq. (\ref{eq:fx}). 
Here, the chemical potential $\mu$ has been chosen in such a way that the Fermi sea of ${\cal H}$ for a fixed number of particles is annihilated by ${\cal H}^{(\pm)}$. 
Since $f(x) = 0$ for $x = 1/2$ (mod $L$), the links between the sites with $x=1$ and $L$ are disconnected by the SSD, resulting in the open boundaries at the edges $x=1$ and $L$.
From the argument for the chiral Hamiltonians above, it follows that the Fermi sea of the original periodic model is an exact eigenstate of the Hamiltonian with SSD. 

So far, we have seen that the Fermi sea of the uniform periodic system is an exact eigenstate of the system with SSD.
For the model with the only nearest neighbor hopping, one can further show that the Fermi sea is indeed the exact {\it ground state} of the SSD Hamiltonian.
The proof is essentially the same as the one presented in Ref.\ \onlinecite{Katsura2011} and we briefly review it here.
We first transform both the original and SSD Hamiltonians into the XY spin chains using the Jordan-Wigner transformation. 
Then, for fixed magnetization (for fixed number of fermions), it can be shown that (i) all of the off-diagonal elements of the Hamiltonian are non-positive, and (ii) the Hamiltonian satisfies the connectivity condition, for both the PBC and SSD cases.
Therefore, the Perron-Frobenius theorem applies and guarantees that the ground state is unique in each case. 
Since the Perron-Frobenius theorem tells us that the ground state is nodeless, i.e., can have only positive components in an appropriate basis, the Fermi sea of the PBC Hamiltonian is the ground state of the SSD Hamiltonian if it has the Fermi sea as an eigenstate.

\subsection{Two and higher dimensions}\label{sec:SSD2d}

In the preceding section, we have shown for the 1d free fermion system that the chiral and SSD Hamiltonians have the Fermi sea of the uniform periodic system as an exact eigenstate.
In this section, we extend the theory to 2d or higher-dimensional systems.
We focus on the single-band model, having single site in each unit cell.
(The multi-band case will be discussed in the end of this section.)
We employ the 2d square lattice as a typical example, but the extension to the other lattices and dimensions is straightforward.

\begin{figure}
\begin{center}
\includegraphics[width=77mm]{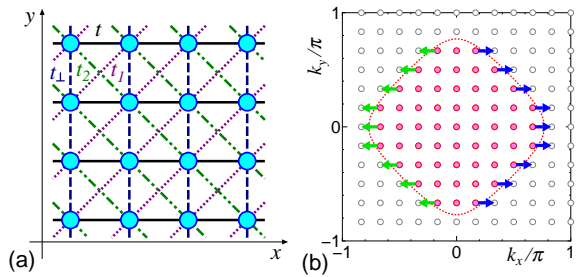}
\caption{
(a) Schematic picture of the square-lattice model (\ref{eq:Ham-Square}). 
(b) Fermi sea of the model (\ref{eq:Ham-Square}) with $t = t_\perp$, $t_1=t_2=0$, and $\mu < 0$. Dotted line shows the Fermi surface. Dark (blue) and light (green) arrows represent respectively the processes allowed by the Pauli principle in the chiral Hamiltonian $\tilde{\cal H}^{(+)}_{\bm \delta}$ and $\tilde{\cal H}^{(-)}_{\bm \delta}$ with ${\bm \delta} = (2\pi/L_x, 0)$.
}
\label{fig:Square}
\end{center}
\end{figure}

Let us consider the tight-binding model in the square lattice,
\begin{eqnarray}
{\cal H} &=& -t \sum_{\bf r} \left[ c^\dagger(x,y) c(x+1,y) + {\rm H.c.} \right]
\nonumber \\
&&-t_\perp \sum_{\bf r} \left[ c^\dagger(x,y) c(x,y+1) + {\rm H.c.} \right]
\nonumber \\
&&-t_1 \sum_{\bf r} \left[ c^\dagger(x,y) c(x+1,y+1) + {\rm H.c.} \right]
\nonumber \\
&&-t_2 \sum_{\bf r} \left[ c^\dagger(x,y) c(x+1,y-1) + {\rm H.c.} \right]
\nonumber \\
&&-\mu \sum_{\bf r} c^\dagger(x,y) c(x,y).
\label{eq:Ham-Square}
\end{eqnarray}
We set the lengths of the system along $x$ and $y$ directions $L_x$ and $L_y$, respectively, and impose the PBC in both directions.
In addition to the nearest-neighbor hoppings $t$ and $t_\perp$ in $x$ and $y$ directions, we have included the diagonal hoppings $t_1$ and $t_2$ for generality.
The schematic picture of the model is shown in Fig.\ \ref{fig:Square} (a).
In momentum space, the Hamiltonian is diagonalized as
\begin{eqnarray}
\tilde{\cal H} = \sum_{\bm k} \epsilon({\bm k}) c^\dagger_{\bm k} c_{\bm k}
\end{eqnarray}
with the dispersion relation
\begin{eqnarray}
\epsilon({\bm k}) &=& 
-2t \cos k_x -2 t_\perp \cos k_y 
\nonumber \\
&&-2t_1 \cos(k_x+k_y) -2t_2 \cos(k_x-k_y) -\mu,
\nonumber \\
&& \label{eq:disp-Square}
\end{eqnarray}
where ${\bm k}=(k_x, k_y)$. 

Similarly to the 1d case, one can introduce the chiral deformation to the system.
The resulting Hamiltonian is
\begin{eqnarray}
&&{\cal H}^{(\pm)}_{\bm \delta}
\nonumber \\
&&= -t \sum_{\bf r} e^{\pm i [\delta_x x + \delta_y (y-\frac{1}{2})]} \left[ c^\dagger(x,y) c(x+1,y) + {\rm H.c.} \right]
\nonumber \\
&&~ -t_\perp \sum_{\bf r} e^{\pm i [\delta_x (x-\frac{1}{2}) + \delta_y y]} \left[ c^\dagger(x,y) c(x,y+1) + {\rm H.c.} \right]
\nonumber \\
&&~ -t_1 \sum_{\bf r} e^{\pm i (\delta_x x + \delta_y y)} \left[ c^\dagger(x,y) c(x+1,y+1) + {\rm H.c.} \right]
\nonumber \\
&&~ -t_2 \sum_{\bf r} e^{\pm i [\delta_x x + \delta_y (y-1)]} \left[ c^\dagger(x,y) c(x+1,y-1) + {\rm H.c.} \right]
\nonumber \\
&&~ -\mu \sum_{\bf r} e^{\pm i [\delta_x (x-\frac{1}{2}) + \delta_y (y-\frac{1}{2})]} c^\dagger(x,y) c(x,y),
\end{eqnarray}
where we consider the cases of ${\bm \delta} = (\delta_x, \delta_y) = (2\pi/L_x, 0), (0, 2\pi/L_y),$ or $(2\pi/L_x, \pm 2\pi/L_y)$. 
From the usual Fourier transformation, the chiral Hamiltonian in ${\bm k}$ space is obtained as
\begin{eqnarray}
\tilde{\cal H}^{(\pm)}_{\bm \delta} = \sum_{\bm k} 
e^{\mp i \frac{1}{2}(\delta_x + \delta_y)} \epsilon({\bm k} \mp {\bm \delta}/2)
c^\dagger_{\bm k} c_{{\bm k} \mp {\bm \delta}}.
\label{eq:Fourier-SquareCD}
\end{eqnarray}
Hence, we find again that the chiral Hamiltonian contains only the processes connecting the sites ${\bm k}$ and ${\bm k} \pm {\bm \delta}$ with an amplitude proportional to the dispersion function Eq.\ (\ref{eq:disp-Square}).
Among them, the processes being active are only those across the Fermi surface, whose amplitudes are of the order of ${\cal O}(1/L_x)$ or ${\cal O}(1/L_y)$.
Therefore, the chiral Hamiltonian does not disturb the Fermi sea very much; The effect of the mixing is of the order of ${\cal O}(1/L_{x,y})$ and vanishes in the limit $L_{x,y} \to \infty$.

Here, the difference from the 1d case should be noted. 
For 1d systems, one can make all the amplitudes of the disturbing processes be exactly zero by tuning the chemical potential $\mu$ as discussed above, and then the Fermi sea can be an exact eigenstate of the chiral Hamiltonian even at finite $L_x$ or $L_y$.
For systems in higher dimensions, on the other hand, it is impossible to achieve $\epsilon({\bm k} \mp {\bm \delta}/2)=0$ for all the processes on the Fermi surface by tuning a single parameter $\mu$, except for some cases in which the Fermi surface takes a special shape.
Therefore, the Fermi sea is an approximate eigenstate of the chiral Hamiltonian and the correspondence becomes exact only asymptotically in the limit $L_{x,y} \to \infty$.
We note that one can achieve the exact correspondence at finite $L_x$ and/or $L_y$ by introducing momentum-dependent chemical potential.
For example, when the chiral deformation in the $x$ direction with ${\bm \delta} = (2\pi/L_x, 0)$ is applied, one may introduce $k_y$-dependent chemical potential $\mu(k_y)$, which leads to long-range hoppings along the $y$ direction in real space, and fine-tune them to realize the zero amplitude for all active processes in the chiral Hamiltonian.

Another notice for higher dimensions concerns the order of taking the infinite-size limits in the $x$ and $y$ directions.
For example, let us consider the case of ${\bf \delta} = (2\pi/L_x, 0)$.
In this case, the amplitudes of the active processes are of the order of ${\cal O}(1/L_x)$ and vanish at $L_x \to \infty$.
However, since the number of these processes is proportional to $L_y$, their effect may be amplified as $L_y$ increases.
In the argument of this section, we take the limit $L_x \to \infty$ first then consider large enough $L_y$.
Taking the limit $L_x=L_y \to \infty$ simultaneously is rather subtle, however, the numerical results presented in Sec. \ref{sec:numerical} suggest that the SSD works well also for the case of $L_x = L_y$.

Let us move to the system with SSD.
By taking an appropriate linear combination of the original and chiral Hamiltonians, one can construct various kinds of the SSD Hamiltonian, which has the form,
\begin{eqnarray}
&& {\cal H}_{\rm SSD} 
\nonumber \\
&& = -t \sum_{\bf r} {\cal F}(x+\frac{1}{2},y) \left[ c^\dagger(x,y) c(x+1,y) + {\rm H.c.} \right]
\nonumber \\
&&~ -t_\perp \sum_{\bf r} {\cal F}(x,y+\frac{1}{2}) \left[ c^\dagger(x,y) c(x,y+1) + {\rm H.c.} \right]
\nonumber \\
&&~ -t_1 \sum_{\bf r} {\cal F}(x+\frac{1}{2},y+\frac{1}{2}) \left[ c^\dagger(x,y) c(x+1,y+1) + {\rm H.c.} \right]
\nonumber \\
&&~ -t_2 \sum_{\bf r} {\cal F}(x+\frac{1}{2},y-\frac{1}{2}) \left[ c^\dagger(x,y) c(x+1,y-1) + {\rm H.c.} \right]
\nonumber \\
&&~ -\mu \sum_{\bf r} {\cal F}(x,y) c^\dagger(x,y) c(x,y).
\label{eq:Ham-SSD}
\end{eqnarray}
The scaling function ${\cal F}(x,y)$ can take various forms; 
for example, ${\cal F}(x,y)=f_x(x)$ for the SSD only in the $x$ direction and 
${\cal F}(x,y)=f_x(x)f_y(y)$ for the SSD in both the $x$ and $y$ directions, 
where $f_x(x)$ [$f_y(y)$] is given by Eq. (\ref{eq:fx}) with $L = L_x$ ($L_y$).
It is clear that these SSD Hamiltonians have the Fermi sea of the original model as an approximate eigenstate. 

We have shown that the chiral and SSD Hamiltonians have the Fermi sea of the uniform periodic system with small disturbance of ${\cal O}(1/L_{x,y})$ as an eigenstate.
However, the argument does not tell us whether the Fermi sea is the ground state of the system with SSD.
As mentioned in the preceding section, for the 1d system, the Perron-Frobenius theorem is applicable and enables us to prove that the Fermi sea is the ground state of the SSD Hamiltonian. 
In contrast, in higher dimensions, it is in general impossible to find a basis in which all of the off-diagonal elements of the Hamiltonian are non-positive, due to fermion signs.
Therefore, the ground state of the SSD Hamiltonian is not necessarily the Fermi sea of the periodic system, and one must examine how the ground state evolves by the SSD using some other methods.
We will show in Sec.\ \ref{sec:numerical} numerically that in two dimension the ground state of the system with SSD is not exactly the Fermi sea but its slight modification accompanied by edge modes localized at boundaries. 

The result above is valid regardless of the boundary conditions in the direction to which the SSD is not applied.
For example, when applying the SSD in the $x$ direction, one can develop the same logic for the system with OBC in the $y$ direction as long as the magnitude of the momentum of the direction is a good quantum number.
The SSD thus changes the boundary condition in a direction from PBC to OBC with the energy deformation, leaving the boundary conditions in the other directions unchanged.

An interesting question is whether one can extend the above argument to the multi-band cases where there are more than one site per unit cell. 
Unfortunately, it turns out that the chiral deformation gives rise to interband hopping terms in the dual lattice (${\bm k}$-space) in these cases. 
Since the Pauli principle does not prohibit the interband hopping processes, the Fermi sea is not expected to be an approximate eigenstate of the SSD Hamiltonian except for some special cases. 
Therefore, we can conclude that the validity of the SSD is basically limited to single-band systems.

\section{Numerical results}\label{sec:numerical}

\begin{figure}
\begin{center}
\includegraphics[width=77mm]{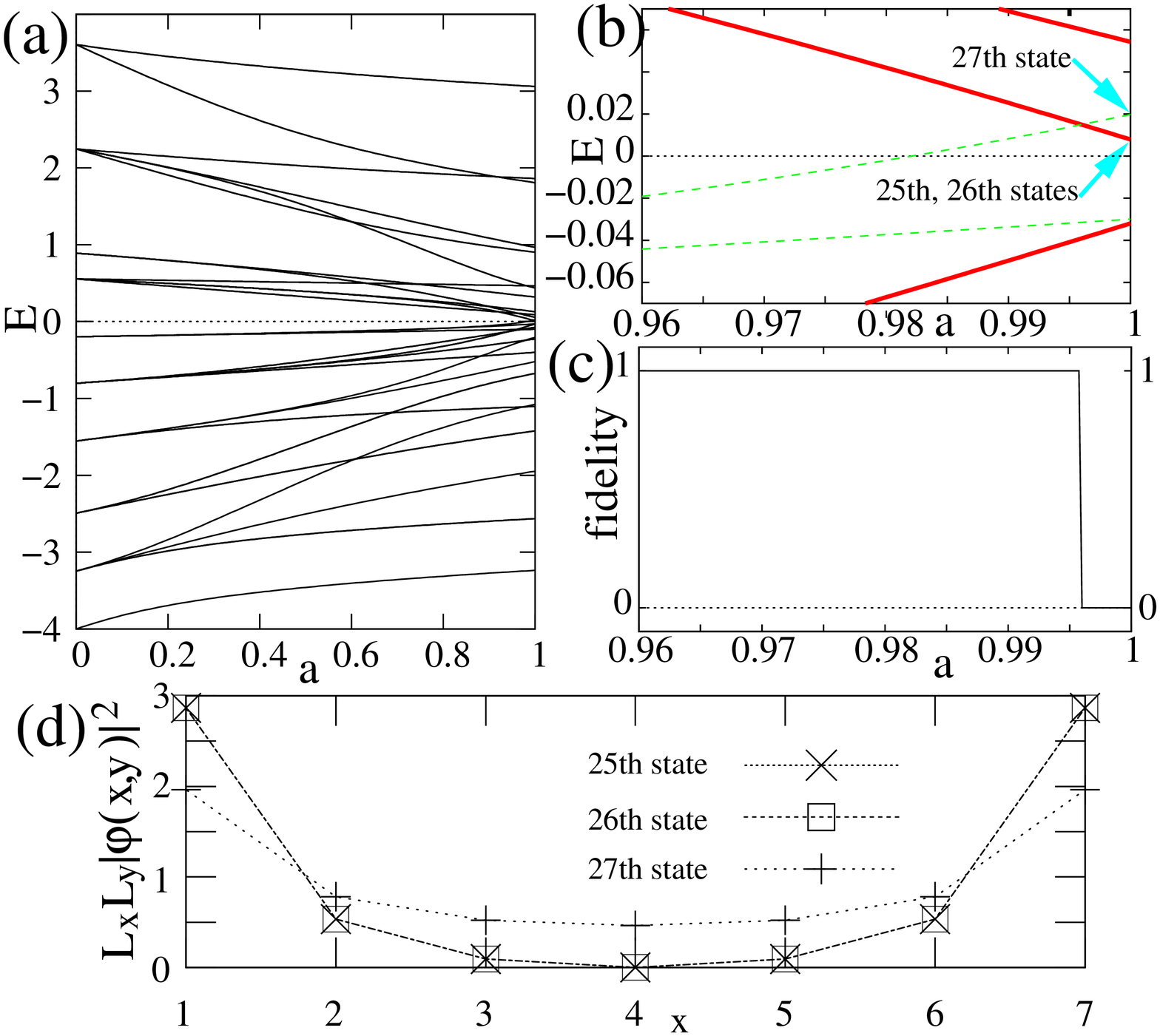}
\caption{
(a) Energy levels of the one-particle eigenstates of ${\cal H}(a)$ for $t=t_\perp=1, t_1=t_2=\mu=0, $ and $L_x=L_y=7$ and (b) its enlarged view around $a=1$.
In (b), thin-dashed and bold lines represent single and doubly-degenerate levels, respectively.
(c) Fidelity between the Fermi sea at $a=0$ and the many-particle ground state of ${\cal H}(a)$ at $a>0$ with $N=25$ particles.
(d) Normalized density profiles $L_xL_y |\varphi_i(x,y)|^2$ of 25, 26, and 27th one-particle eigenstates at $a=1$ as a function of $x$.
Note that the profiles are independent of $y$ because of the PBC imposed in the $y$ direction.
}
\label{fig:PBC-SSD1}
\end{center}
\end{figure}

\begin{figure}
\begin{center}
\includegraphics[width=65mm]{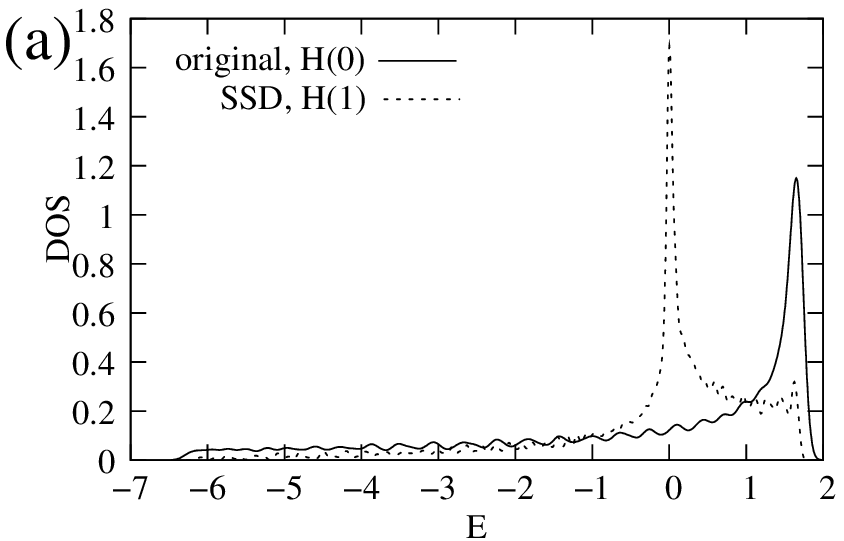}
\includegraphics[width=65mm]{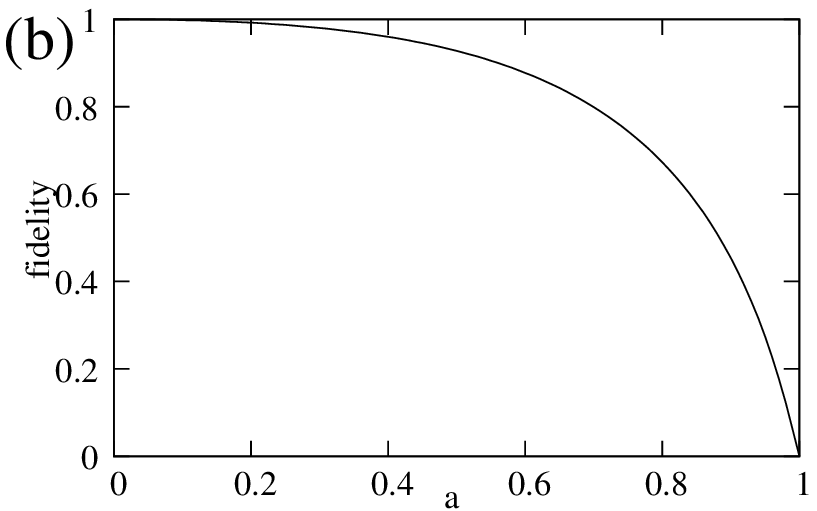}
\includegraphics[width=65mm]{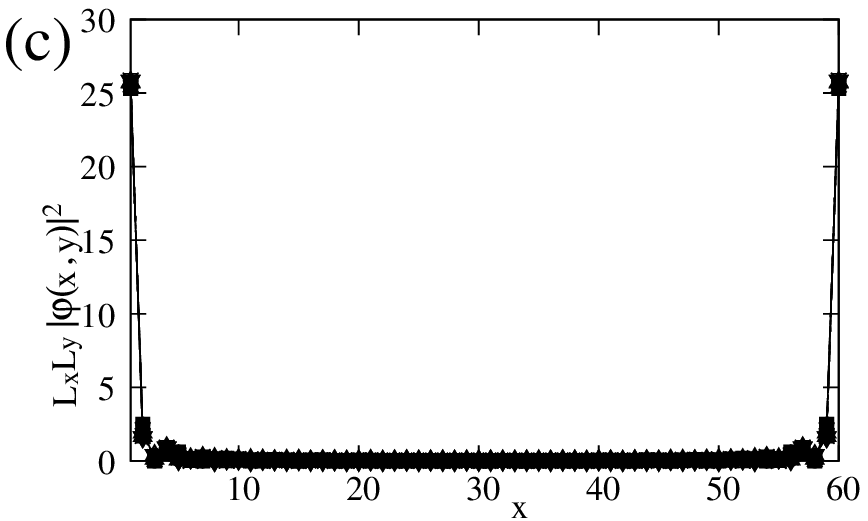}
\includegraphics[width=65mm]{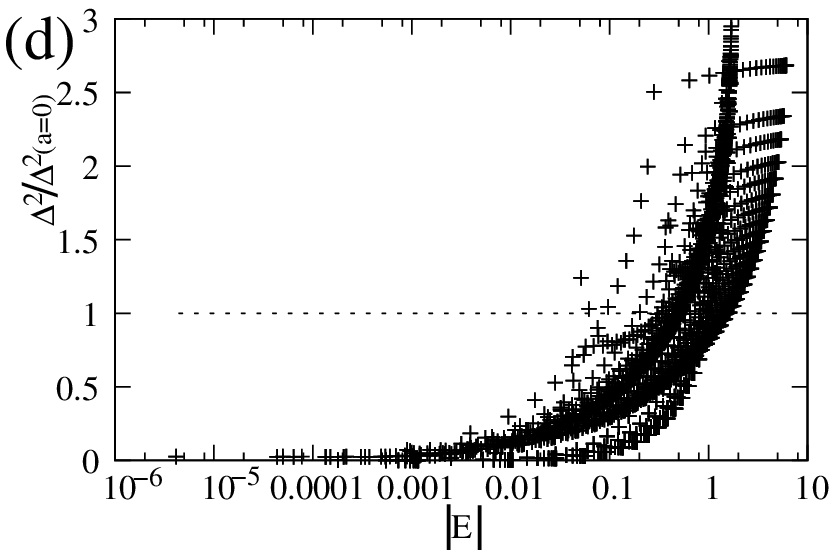}
\caption{
Numerical results for the square-lattice model with diagonal hoppings, $t=t_\perp=1, t_1=t_2=0.5, \mu=0.3,$ and $L_x=L_y=60$;
(a) Density of state, where we use the Gaussian broadening with $\sigma=0.5$.
(b) Fidelity between the Fermi sea at $a=0$ and the many-particle ground state at $a>0$ with fixed particle number.
(c) Normalized density profiles $L_x L_y |\varphi_i(x,y)|^2$ at $a=1$. The results for eleven states with the smallest $|E_i|$ are plotted.
(d) $\Delta_i^2$ as a function of $|E_i|$.
The data are normalized by the value for the uniform periodic system, $\Delta_i^2(a=0)={(L_x-1)(L_x-2)/12}$. 
The horizontal dotted line is a guide for the eye.
}
\label{fig:PBC-SSD2}
\end{center}
\end{figure}

In this section, we discuss how the ground state changes when we apply the SSD to the 2d system under PBC. 
For the purpose, we introduce the model connecting the uniform system and the system with SSD,
\begin{eqnarray}
{\cal H}(a) = (1-a){\cal H} + a {\cal H}_{\rm SSD},
\end{eqnarray}
where we consider the case that the SSD is applied in $x$ direction.
Therefore, ${\cal H}(a)$ is given by the r.h.s. of Eq.\ (\ref{eq:Ham-SSD}) with replacing the scaling function ${\cal F}(x,y)$ by 
\begin{eqnarray}
f(x; a) = (1-a) + a f_x(x).
\label{eq:fx-a}
\end{eqnarray}
Note that $a=0$ corresponds to the uniform system with PBC while $a=1$ gives the system with SSD.

First, we consider the simple square lattice with $L_x=L_y=L$, $t=t_\perp=1$ and $t_1=t_2=\mu=0$, where the Fermi surface has the diamond shape for infinite $L$.
For finite and odd $L$, the amplitude $\epsilon(k_x \mp \frac{\delta_x}{2}, k_y)$ in the chiral Hamiltonian, Eq.(\ref{eq:Fourier-SquareCD}), becomes exactly zero for the processes across the Fermi surface, therefore the Fermi sea of the uniform model with $a=0$ is an exact eigenstate for arbitrary $a$.
For even $L$, the Fermi sea at $a=0$ becomes an exact eigenstate for arbitrary $a$ when the boundary condition in the $y$ direction is anti-periodic. 
We again emphasize that an exact eigenstate does not mean the ground state.
In fact, the Fermi sea at $a=0$ is, in general, not the ground state but an excited state for $a=1$ due to the level crossing, as shown below.

Figures\ \ref{fig:PBC-SSD1}(a) and (b) show the energy levels of the one-particle eigenstates of ${\cal H}(a)$ for $L=7$ with PBC in the $y$ direction.
With increasing $a$, many energy levels are approaching to $E=0$.
At $a=0.982$, a state crosses the Fermi energy $E=0$, leading to the change of the particle number of the ground state filling the negative energy.
Furthermore, at $a=a_c=0.996$, the level crossing occurs.
Therefore, even if the particle number $N$ is fixed at that of $a=0$, the ground state changes abruptly at $a = a_c$ and the Fermi sea at $a=0$ is no longer the ground state for $a>a_c$.
To elucidate it, we plot in Fig.~\ref{fig:PBC-SSD1}(c) the fidelity, i.e., wave-function overlap, between the Fermi sea at $a=0$ and the ground state of ${\cal H}(a)$ for fixed $N=25$ particles.
The fidelity jumps from 1 to 0 at the level crossing point $a=a_c$.
We also confirmed numerically that the Hamiltonian ${\cal H}(a)$ has the Fermi sea at $a=0$ as an exact, first-excited eigenstate for $a>a_c$.

Here, it is non-trivial that the ground state at $a=0$ defined by the Slater determinant of the negative-energy states is the exact eigenstate of ${\cal H}(a)$ in spite that the one-particle eigenstates are different.
While the local density of one-particle states at $a=0$, which are a simple plane wave, does not depend on the spatial position $(x,y)$, the density for $a > 0$ depends on $x$ due to the energy deformation in $x$ direction.
In Fig.~\ref{fig:PBC-SSD1}(d), we show the normalized density profiles $L_xL_y |\varphi_i(x,y)|^2$ of three low-energy one-particle states $\varphi_i(x,y)$ at $a=1$, which states undergo the level crossing at $a=a_c$.
As shown in the figure, those states are localized at the boundaries at $x=1$ and $L$.
The result implies that the difference between the Fermi sea of the uniform system ${\cal H}={\cal H}(0)$ and the ground state of ${\cal H}_{\rm SSD}={\cal H}(1)$ manifests itself in the boundary region.

Next, we show that the above features of SSD are commonly seen in the more generic case $t=t_\perp=1$, $t_1=t_2=0.5$, and $\mu=0.3$, where the Fermi sea of ${\cal H}(0)$ is not the exact ground state nor even an exact eigenstate for $a > 0$ due to small disturbance of ${\cal O}(1/L_x)$ by the chiral Hamiltonians.
Figure\ \ref{fig:PBC-SSD2}(a) shows the density of states at $a=0$ and $a=1$ 
for the systems with $L_x = L_y = 60$.
The prominent peak at $E=0$ for $a=1$ indicates that many levels approach $E=0$ after the SSD, as in Fig.~\ref{fig:PBC-SSD1}(a). 

Figure\ \ref{fig:PBC-SSD2}(b) shows the fidelity between the Fermi sea at $a=0$ and the ground state at $a > 0$ with the fixed particle number.
Due to the small disturbance mentioned above, the fidelity is not exactly one even for small $a$ in contrast to Fig.~\ref{fig:PBC-SSD1}(b).
In the usual perturbation theory via small $a$, the fidelity is linear in $a$ and its gradient is determined by the expectation value of the perturbative Hamiltonian in the unperturbed ground state.
Therefore, the fact that the fidelity at small $a$ remains almost unity means that the expectation value of ${\cal H}^{(\pm)}$ is very small.
The fidelity exhibits a continuous decrease from unity at $a=0$ to almost zero at $a = a_c = 0.9991$, followed by a small jump to zero at $a=a_c$ [which is almost invisible in Fig.~\ref{fig:PBC-SSD2}(b)].
We have observed that the finite jump of the fidelity, which is due to the level-crossing of the ground state and excited state, generally occurs at $a = a_c \sim 1$.

We present in Fig.\ \ref{fig:PBC-SSD2}(c) the normalized density profiles of the one-particle states with smallest $|E_i|$ at $a=1$, which are expected to be the most responsible for the change in the ground-state properties.
We note that the figure plots eleven states locating closest to the Fermi energy.
It is clear that the profiles of those states are almost identical and strongly localized around the open boundaries $x=1$ and $L_x$, in marked contrast to the plane waves for $a=0$ whose profiles are a constant.
We note that the decay of the profiles from the open edges seems to be algebraic, while determining the detailed form of the decay function as well as the decay exponent requires a more systematic analysis, which is left for future studies.

To further clarify the edge-localized nature of the low-energy eigenstates, we introduce a measure of the localization, 
\begin{eqnarray}
\Delta^2_i = {\sum_{x,y}|\varphi_i(x,y)|^2  \mbox{min}\left[ (x-1)^2,(x-L_x)^2 \right] }.
\end{eqnarray}
Note that $\Delta^2_i$ is small for localized states while it becomes large for the states extended to the bulk.
For the uniform system ${\cal H}={\cal H}(0)$, $\Delta_i^2$ is easily obtained to be $\Delta_i^2={(L_x-1)(L_x-2)/12}$ for even $L_x$ since $|\varphi_i(x,y)|^2=1/(L_xL_y)$ for all eigenstates.
For $a=1$, we obtain numerically all of the eigen-energies and eigen-wavefunctions $(E_i, \varphi_i)$ and calculate $\Delta^2_i$.
Figure\ \ref{fig:PBC-SSD2}(d) indicates that all the eigenstates with small $|E_i|$ have very small $\Delta^2_i$.\cite{extended_state}
This means that the states around the Fermi level, which get closer and converge to $E=0$ as $a$ increases from $0$ to $1$, are strongly localized around the open edges.
The localization of low-energy states can be naturally understood from the fact that the scaling function of the SSD, $f_x(x)$, is small around the open edges, therefore the excitations around the edges can appear with an extremely low energy cost.

\begin{figure}
\begin{center}
\includegraphics[width=65mm]{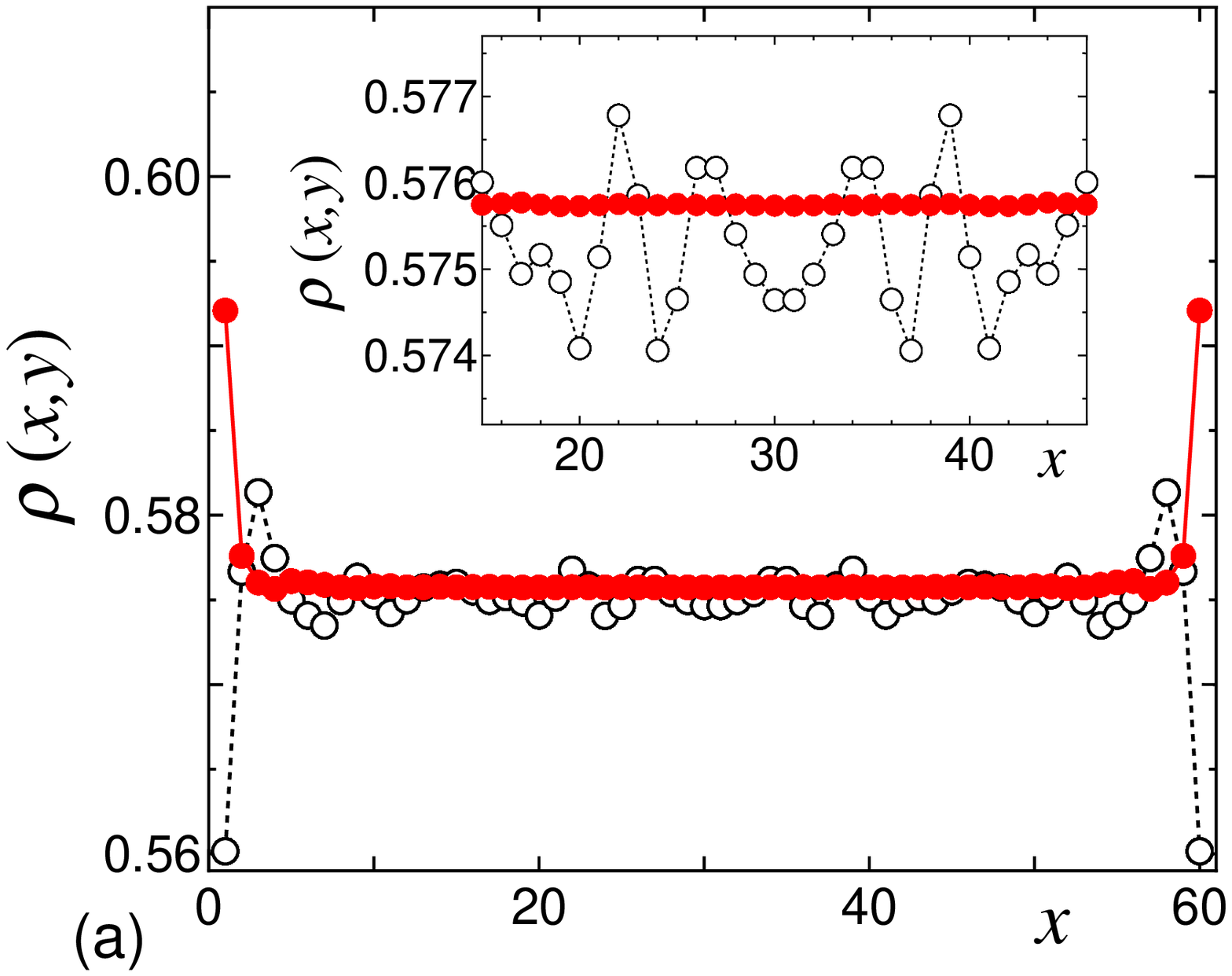}
\includegraphics[width=65mm]{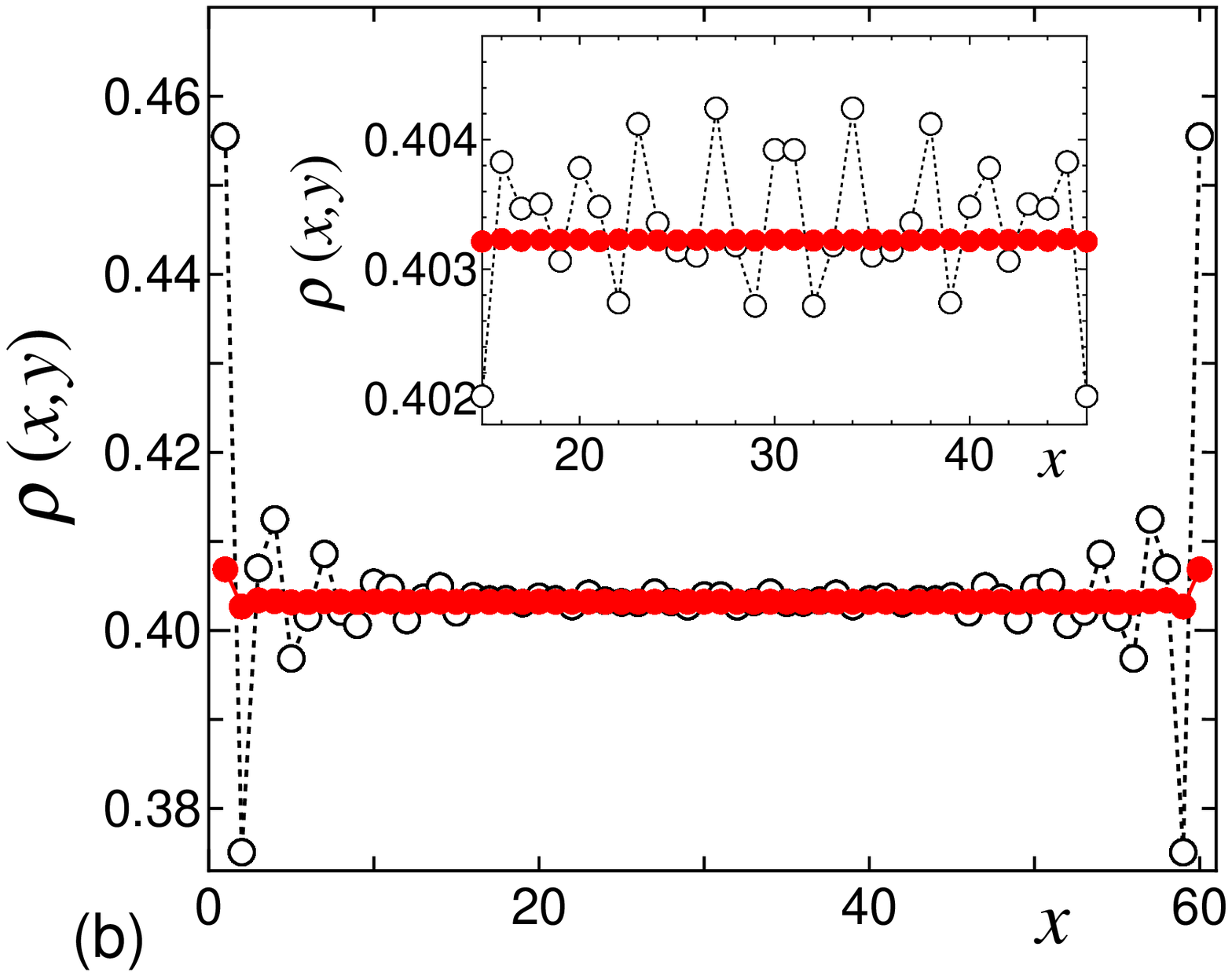}
\includegraphics[width=65mm]{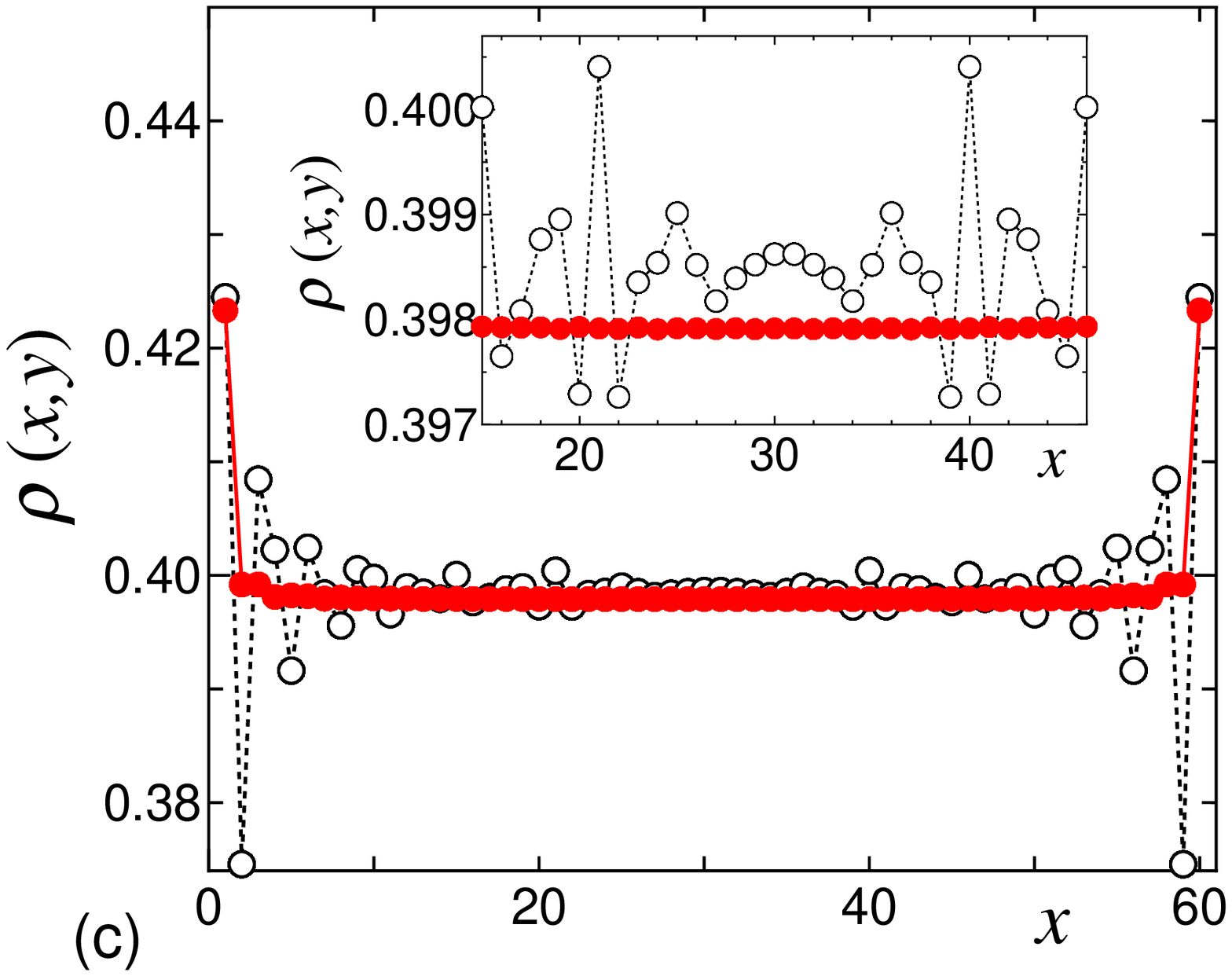}
\caption{
Density profiles $\rho(x,y)$ in the ground state of 
(a) the square-lattice model, $t=t_\perp=1$, $t_1=t_2=0$, and $\mu = 0.3$, 
(b) the square-lattice model with diagonal hoppings, $t=t_\perp=1$, $t_1=t_2=0.5$, and $\mu = 0.3$, and 
(c)  the triangular-lattice model, $t=t_\perp=t_1=1$, $t_2=0$, and $\mu = 0$.
The system size is $L_x=L_y=60$.
Solid and open circles respectively show the results for the system with SSD and the uniform system with OBC in the $x$ direction.
The PBC is imposed for the $y$ direction, therefore $\rho(x,y)$ is independent of $y$.
Insets present the enlarged figure of the same data for $15 \le x \le 46$.
}
\label{fig:rho-SSD}
\end{center}
\end{figure}

\begin{figure}
\begin{center}
\includegraphics[width=75mm]{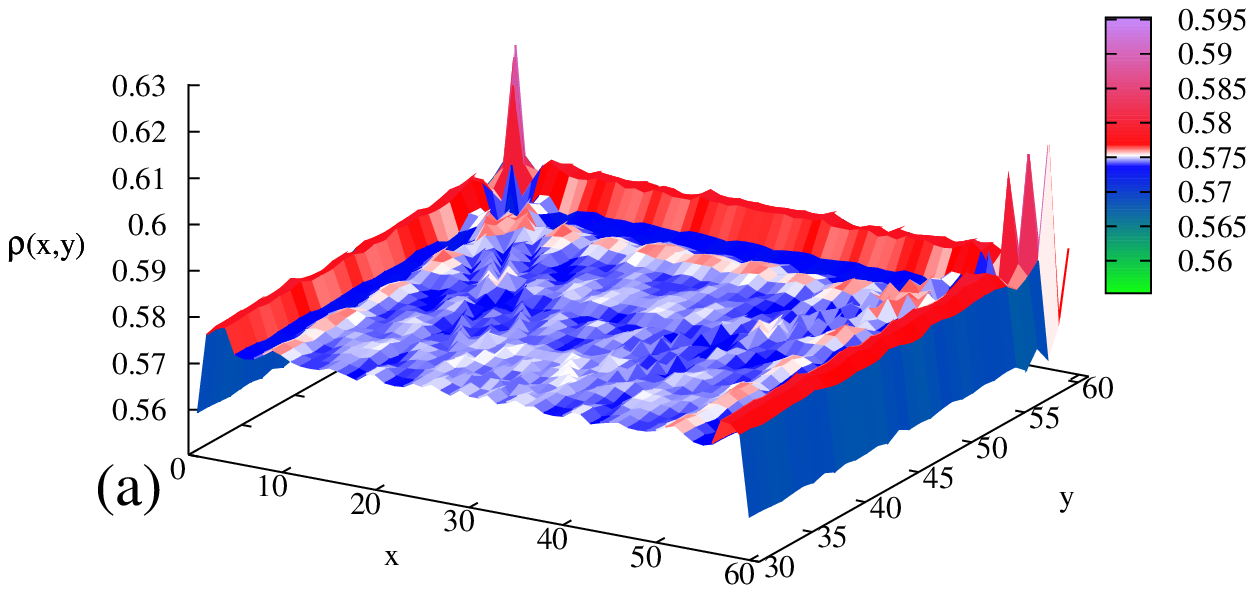}
\includegraphics[width=75mm]{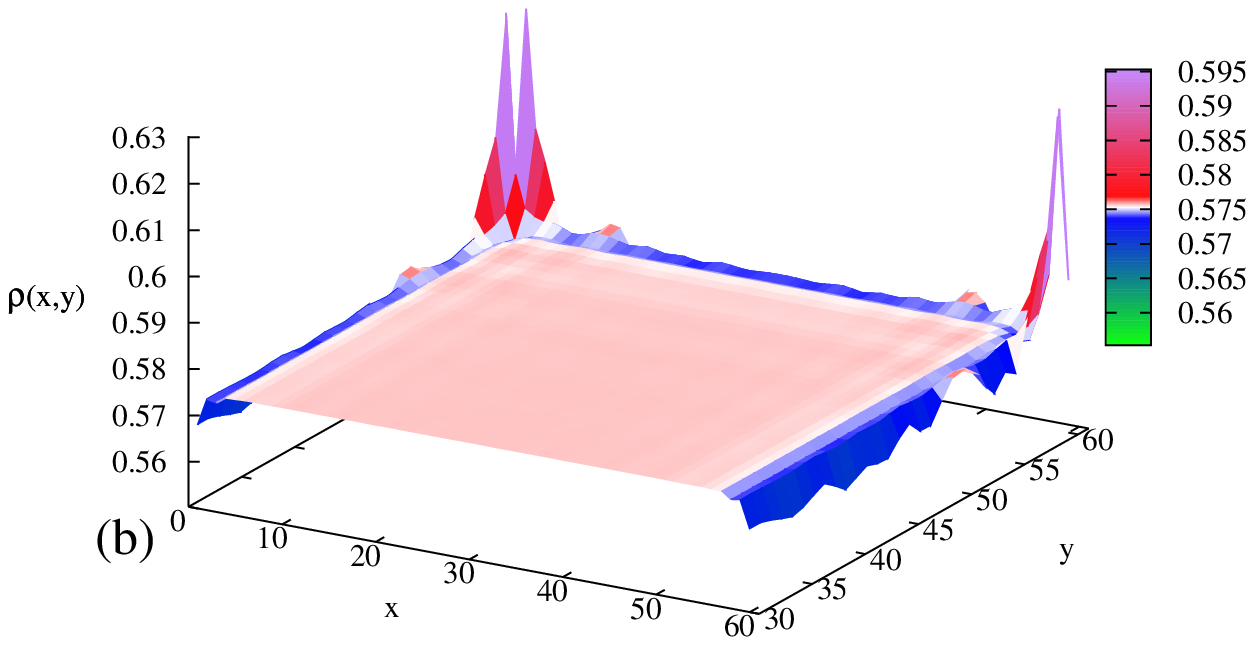}
\caption{
Density profiles $\rho(x,y)$ in the ground state of the square-lattice model, $t=t_\perp=1$, $t_1=t_2=0$, $\mu = 0.3$, and $L_x=L_y=60$; 
(a) the result for the OBC imposed in both $x$ and $y$ directions, and 
(b) for the SSD in both the directions.
The data for $30 \le y \le 60$ are plotted while those for $1 \le y < 30$ are symmetric with respect to $y = (L_y+1)/2$.
}
\label{fig:rho-SSDxy}
\end{center}
\end{figure}

Finally, we investigate how the SSD affects properties of the many-body ground state.
In Fig.\ \ref{fig:rho-SSD}, we show the density profiles 
\begin{eqnarray}
\rho(x,y) = \langle c^\dagger(x,y) c(x,y) \rangle 
= \sum_{i~(E_i<0)} |\varphi_i(x,y)|^2
\end{eqnarray}
in the ground state of the system with SSD in $x$ direction [Eq.\ (\ref{eq:Ham-SSD}) with ${\cal F}(x,y)=f_x(x)$] for several typical parameter sets.
The density $\rho(x,y)$ in the ground state of the uniform system with OBC in $x$ direction is also shown for a comparison.
Remarkably, the density profile $\rho(x,y)$ in the systems with SSD is almost a constant in the bulk and the Friedel oscillations pronounced in the uniform system with OBC are strongly suppressed.
The steep change in $\rho(x,y)$ from the constant is found only in the vicinity of the open edges.
The results indicate that the SSD preserves the translationally-invariant nature of the Fermi sea of the periodic ground state in the bulk and the deviation appears only around the open edges.

Figure\ \ref{fig:rho-SSDxy} shows the density profiles in the square-lattice system with SSD in both $x$ and $y$ directions, whose Hamiltonian is given by Eq.\ (\ref{eq:Ham-SSD}) with ${\cal F}(x,y)=f_x(x) f_y(y)$.
The result for the system with OBC in both the directions is also shown.
Again, we find that $\rho(x,y)$ is nearly a constant in the bulk of the system with SSD and the Friedel oscillations, which are conspicuous in the system with OBC, are removed almost completely.
In the system with SSD, the deviation in $\rho(x,y)$ from the constant is sizable only around the open edges as well as the corners.
We have observed essentially the same results for the square-lattice model with diagonal hoppings and triangular-lattice model.
The completely flat profiles strongly suggests that the ground state of the systems with SSD is very close to the translationally-invariant ground state of the uniform periodic system superposed by the small disturbance of the edge states.
The SSD thereby realizes the topology change of the system from a torus to a cylinder, and then, to a rectangle with keeping the ground-state properties in the bulk unchanged.

\section{Concluding remarks}\label{sec:conclusion}

We have studied effects of the SSD on the free fermion systems.
We have proposed a simple theory of the mechanism of the SSD to realize the energy deformation with keeping the ground state of the periodic system unchanged.
For a certain class of 1d systems, it is shown that the Fermi sea of the uniform system remains an exact ground state of the system with SSD.
For 2d systems, it is numerically found that the SSD realizes almost flat density profiles in the bulk, suggesting that the ground state properties in the bulk are translationally invariant in spite that the system has open edges.
The ground state of the two- and higher-dimensional systems with SSD can be regarded as a Fermi sea of the uniform periodic system superposed by edge-localized states.

As we have shown, in one dimension, the Fermi sea is not only an exact eigenstate but also the ground state of the Hamiltonian with SSD if the chemical potential $\mu$ is chosen so that $\epsilon(k \mp \delta/2)=0$ for the processes in the SSD Hamiltonian allowed by the Pauli principle. 
One can explicitly confirm this correspondence in a class of exactly solvable spin chains.\cite{Katsura_in_prep} 
It is worth mentioning that the Hamiltonian with SSD is distinct from the inhomogeneous integrable models obtained from the quantum inverse scattering method and the algebraic Bethe ansatz.\cite{Schmitteckert_Schwab95, Moliner_Schmitteckert10}
From the point of view of conformal field theory, the chiral Hamiltonian has a definite meaning, which accounts for the correspondence between the PBC and SSD Hamiltonians in a wide class of 1d critical systems. 
A detailed discussion will also be given in a subsequent paper.\cite{Katsura_in_prep}

Our results demonstrate that the SSD works extremely well for free fermions in any dimensions.
A natural question to ask is whether the results generalize to systems with interactions. 
In one dimension, it has been confirmed numerically for several strongly-correlated fermion\cite{GendiarDLN2011,ShibataH2011} and spin\cite{HikiharaN2011} systems that the SSD works efficiently to realize the translationally-invariant ground state, and the results are well founded by a field-theoretical analysis.\cite{Katsura_in_prep}
Extending the analysis to multi-leg ladder systems,\cite{LinBF1997} and then, to 2d systems must be an intriguing problem.
It would also be interesting to explore the generalizations of other deformations such as the hyperbolic,\cite{Ueda-Nishino08, Ueda-Nakano08, Ueda-Nakano10} exponential,\cite{Okunishi-Nishino10} and sinusoidal\cite{GendiarDLN2011} deformations, to higher-dimensional and/or interacting systems. 

An experimental study of the SSD in real systems should also be a challenge of great interest.
Considering the fact that hopping strength and coupling parameters can be tuned in systems of ultracold atoms in optical lattices as demonstrated in Refs.\ \onlinecite{Roati2008,Bakr2009,Yamazaki2010}, they can be promising candidates for the realization of the system with SSD studied in our work.

\acknowledgments

We thank Tomotoshi Nishino for stimulating discussions. 
I.M. was supported in part by Grant-in-Aid for Young Scientists (B) (20740214).
H.K. was supported in part by Grant-in-Aid for Young Scientists (B) (23740298). T.H. was supported in part by the MEXT and JSPS, Japan through Grand Nos.\ 21740277 and 22014016.


\begin{thebibliography}{99}
\bibitem{extended_state}
To be precise, in some rare cases that there is a quantized momentum $(k_x, k_y)$ being very close to the Fermi surface at $a=0$, there can be a one-particle state for $a=1$ which is extended to the bulk and has small $|E_i|$. 
However, such a state is present already at $a=0$ and does not concern the change of the ground-state properties induced by the SSD.
We have selected the parameters in Fig.\ \ref{fig:PBC-SSD2} to avoid such a rare event.
\end{thebibliography}
\end{document}